\documentclass[14pt]{article}
\usepackage{srcltx}
\usepackage[utf8]{inputenc}
\usepackage[english]{babel}
\usepackage{graphicx}
\usepackage{subcaption}
\usepackage{amsmath,amsfonts,amssymb,amsthm,mathtools} 
\usepackage{booktabs}
\usepackage{xcolor}
\begin{document}
\title{ Uniform models of neutron and quark
(strange) stars in General Relativity}

\author{%
G.\,S.\,Bisnovatyi-Kogan\thanks{Space Research Institute, Russian Academy of Sciences, Moscow, Russia; National Research Nuclear University MEPhI (Moscow Engineering Physics Institute), Moscow, Russia; Moscow Institute of Physics and Technology, Dolgoprudny, Moscow oblast, Russia},
E.\,A.\,Patraman\thanks{Space Research Institute, Russian Academy of Sciences, Moscow, Russia; Moscow Institute of Physics and Technology, Dolgoprudny, Moscow oblast, Russia}
}

\maketitle

\begin{abstract}
  Models of neutron and strange stars are studied within the
  approximation of a uniform density distribution. A universal
  algebraic equation, valid for any equation of state, is used to
  estimate the stellar mass at a given density without resorting to
  the numerical integration of differential equations. Equations of
  state for neutron stars include both a degenerate neutron gas and
  more realistic models, such as those employed by  Malone, 
  Johnson and Bethe \cite{bethe}. Homogeneous strange star models based on the
  quark bag model equation of state admit simple analytical
  solutions. The approximate solutions presented in this work differ
  from the exact results obtained by numerical integration of the
  structure equations by no more than $\sim 20\%$.

The formation of strange stars is examined as a function of the deconfinement boundary (DB), at which quarks become deconfined. Existing experimental data indicate that matter reaches extremely high densities in the vicinity of the DB. This places strong constraints on the maximum mass of strange stars and disfavors their formation at the final stages of stellar evolution, since the limiting mass of neutron stars is substantially higher and corresponds to significantly lower matter densities.
\end{abstract}

\section{Introduction}
The first conclusion regarding the existence of an upper mass limit for cold stars supported by the pressure of degenerate electrons was reached by Stoner \cite{stoner30}, who considered a uniform-density white dwarf model. Generalizing earlier studies of degenerate electron pressure \cite{fowler26, frenkel28, azp29} to the regime of ultra-relativistic degeneracy at high densities, he obtained a limiting mass of $M_{wd} = 1.1\,M_{\odot}$ for $\mu_e = 2.5$. 

To refine the maximum mass of observed white dwarfs (WD), Chandrasekhar \cite{chandra31}, following Stoner \cite{stoner30}, adopted $\mu_e = 2.5$ but modeled the stellar structure using the Emden polytropic solution with index $n = 3$, obtaining $M_{wd} = 0.933\,M_{\odot}$. However, both stellar evolution theory and observations indicate that most white dwarfs are composed of a mixture of carbon ($^{12}$C) and oxygen ($^{16}$O), for which $\mu_e = 2$ \cite{shatzman}, yielding a limiting mass of $M_{wd} = 1.46\,M_{\odot}$. This realistic value was first obtained by Landau \cite{landau32}, independently of Stoner and Chandrasekhar. The pioneering role of D. Stoner in formulating the idea of a limiting mass of WD was described in detaileds in Refs. \cite{mn2008,et2011}.

In this paper, we construct approximate models of cold neutron and strange stars of arbitrary mass under the assumption of a uniform density distribution, using algebraic equations derived within the framework of general relativity (GR). Within this approach, all results—including the limiting masses of neutron stars—are obtained analytically from algebraic equations derived in \cite{bkp23} (see also \cite{nch1973}), and apply to arbitrary equations of state. For the widely used quark bag model equation of state \cite{wit}, a particularly simple analytical solution is obtained.

Uniform models of neutron stars had been investigated also in the frame of extended GR gravity \cite{pappas}.

The equilibrium models of uniform spherical  bodies of non-compressible fluid in General Relativity are described in textbooks, see e.g. \cite{w72,mtw73,st83}.
We consider here different stars, with realistic equations of state, and use the variation  principle 
for  constructing  models, with a prescribed  uniform density distribution \cite{stoner30,zn71}. We obtain algebraic equation from variation approach in GR, which determine integral properties of stars: mass, radius, density, pressure, entropy. From these equations we calculate  approximate values of these parameters, at different density,with a fixed mass, up to critical states, where stars loose their stability. 
\color{black}

\section{Uniform-density neutron stars}

To construct realistic models of neutron stars (NSs) and strange stars (SSs), it is necessary to employ general relativity, since the gravitational potential reaches values of order tenths of $c^2$, and the NS radius is only a few gravitational radii, $R_g = 2GM/c^2$. Models of non-rotating NSs are constructed using a Schwarzschild-type metric  \cite{zn71,tolman,oveq,lltp}

\begin{eqnarray}
\label{eq19}
ds^2=e^{\nu(r)}c^2dt^2-e^{\lambda(r)}dr^2-r^2(d\theta^2+\sin^2\theta d\phi^2),\quad 
e^{\nu} = e^{-\lambda}=1-\frac{2Gm}{rc^2},\qquad \\
e^{\nu} = \Big( 1 - \frac{2Gm}{rc^2} \Big) \exp\Bigg[ \int_0^{P(r)} \frac{2 dP}{P + \rho(P)} \Bigg] . \qquad \qquad \nonumber 
\end{eqnarray}
where

\begin{eqnarray}
\label{eq20}
m(r)=4\pi\int_0^r \rho r^2 dr, \quad M=m(R), \quad \rho=\rho_0\left(1+\frac{E_T}{c^2}\right).
\end{eqnarray}
In general relativity, the total density of matter $\rho$ is used, which includes the rest-mass energy density $\rho_0$ and the internal energy $E_T$. The mass $M$ is the total gravitating mass, which includes the gravitational binding energy; therefore, the total energy of an NS is $E = M c^2$, and the total energy $e$ contained within a radius $r$ is $e(r) = m(r)c^2$. From the equations of general relativity, it follows that the baryon density $n$ is related to the number of baryons (or quarks) $f(r)$ contained within radius $r$ as \cite{bk89}

\begin{eqnarray}
\label{eq21}
f(r)= 4\pi\int_0^r n\, r^2\left(1-\frac{2Ge}{c^2 r}\right)^{-1/2} dr, \nonumber \\ 
\quad N=f(R), \quad \rho_0=n\,m_u, \quad M_0=N\,m_u,
\end{eqnarray}
where $M_0$ is the baryonic rest mass of the NS,  $m_u$ is the mass unit, equal to 
$\frac{1}{12} m_{^{12}C}$; $n$ is the number density. 

\begin{figure}[h]
\begin{center}
\includegraphics[width=0.7\linewidth]{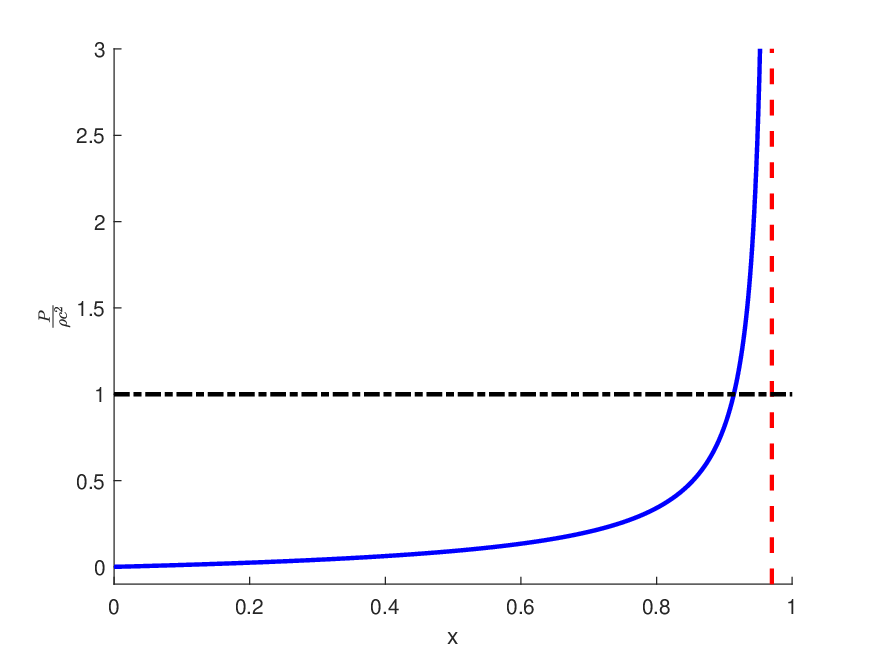}
\end{center}
\caption{(Solid line) Dependence of $\frac{P}{\rho c^2}$ on the parameter $x$, according to equation~(\ref{eq7}) . The abscissa of the vertical dashed line at $x = x_l = 0.9849$, corresponds to the zero of the denominator of equation~(\ref{eq7}). The horizontal dash-dotted line $\frac{P}{\rho c^2}=1$ separates the physically admissible region from the upper region, where the principle of causality, requiring sound speed $v_s\le c$, is violated, see \cite{zeld62}.}
\label{1}
\end{figure}

For a uniform star, the quantities $n$, $\rho_0$, $E_T$, and $\rho$ do not depend on radius, and the integral in expression \eqref{eq21} can be evaluated analytically \cite{ryzhik}

\begin{eqnarray}
\label{eq22}
f(r)= \frac{2\pi n}{(D)^{3/2}}\left[\sin^{-1}(r\sqrt{D})-r\sqrt{D}\sqrt{1-r^2D}\right], \quad D=\frac{8\pi \rho G}{3c^2}.
\end{eqnarray}
To relate the NS radius $R$ to the density $\rho$, it is necessary to find the extremum of the function $M(\rho_0)$ at a fixed number of baryons in the star, $N$, which are given by

\begin{eqnarray}
\label{eq23}
M=\frac{4\pi}{3}\rho R^3=\frac{4\pi}{3 c^2}\rho_0(c^2+E_T) R^3, \qquad
N= \frac{2\pi \rho_0}{m_u D^{3/2}}\left[\sin^{-1}(R\sqrt{D})-R\sqrt{D}\sqrt{1-R^2D}\right].
\end{eqnarray}
The differentials $dM$ and $dN$ are linear combinations of the differentials $dR$ and $d\rho_0$. For a fixed number of baryons, $dN = 0$, this yields a linear relation between $dR$ and $d\rho_0$, and thus determines the dependence $M(\rho_0)$. The equilibrium condition

\begin{equation}
\label{eq30}
\frac{dM}{d\rho_0}=0
\end{equation}
leads to the following algebraic equation for the equilibrium of a uniform barotropic NS or SS in general relativity \cite{bkp23}

\begin{equation}
\label{eq7}
    \frac{P}{\rho c^2} = \frac{\Phi_0(x)}{\Phi_1(x)}.
\end{equation}
Here,

\begin{eqnarray}
\label{eq43b}
\Phi_0(x)= 2 x^3-3\sqrt{1-x^2}(\,\sin^{-1}x-x\sqrt{1-x^2})=
x(3-x^2)-3\sqrt{1-x^2}\sin^{-1}x ,\\
\Phi_1(x)= -2x^3+9\sqrt{1-x^2}(\sin^{-1}{x}-x\sqrt{1-x^2})=
 x(7x^2-9)+9 \sqrt{1-x^2}\sin^{-1}x . \nonumber
\end{eqnarray}

\begin{eqnarray}
\label{eq41}
x=R\sqrt{D}=\sqrt{\frac{R_g}{R}}, \qquad R_g=\frac{2 GM}{c^2}.
\end{eqnarray}
Figure~\ref{1} shows the dependence of $\frac{P}{\rho c^2}$ on the parameter $x$, as determined by equation~(\ref{eq7}). When the denominator on the right-hand side of equation~(\ref{eq7}) approaches zero, the ratio $\frac{P}{\rho c^2}$ and the density $\rho$ diverge. The abscissa of the vertical dashed line, $x = x_l$, in Fig.~\ref{1} corresponds to the zero of the denominator of equation~(\ref{eq7}), with $x_l = 0.9849$. The horizontal dash-dotted line $\frac{P}{\rho c^2}=1$ separates the physically admissible region from the upper region, in which the principle of causality is violated, since the speed of sound in matter cannot exceed the speed of light.

\noindent

\section{Construction of the equilibrium curve $M(\rho)$ for a uniform density model}

The dependences $M(\rho)$ or $M(\rho_0)$ and $M(R)$ can be constructed using the following procedure.

1. Select an equation of state $P(\rho)$ or $P(\rho_0)$ given by analytical expressions or tabulated data.

2. Specify the ratio of the gravitational radius of the star to its physical radius:
$x^2 = \frac{R_g}{R}$.

3. Determine the ratio of pressure to density for this value of $x$.

4. Find the corresponding value of the density from the equation of state.

5. Determine the radius of the model from the definition of $x$ in the form
$R^2 = \frac{3c^2 x^2}{8\pi G\rho}$.

6. Finally, calculate the mass of the uniform-density model,
$M = \frac{4\pi}{3} \rho R^3$.
\newline

\begin{figure}[!h]
\begin{center}
\includegraphics[width=0.65\linewidth]{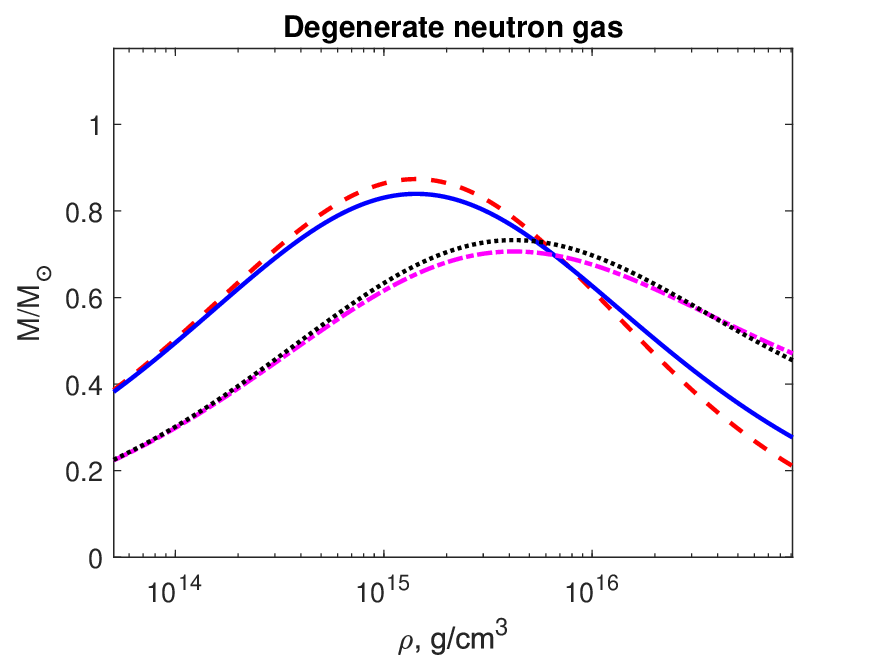}
\end{center}
\caption{Degenerate neutron gas.  Dependence $M(\rho)$ for the uniform density (solid line) and for the
exact models (dash-dotted line). Dependence $M_0(\rho)$ for the uniform-density (dashed line), and for the
exact model (dotted line).}
\label{2}
\end{figure}

\begin{figure}[h!]
\begin{center}
\includegraphics[width=0.65\linewidth]{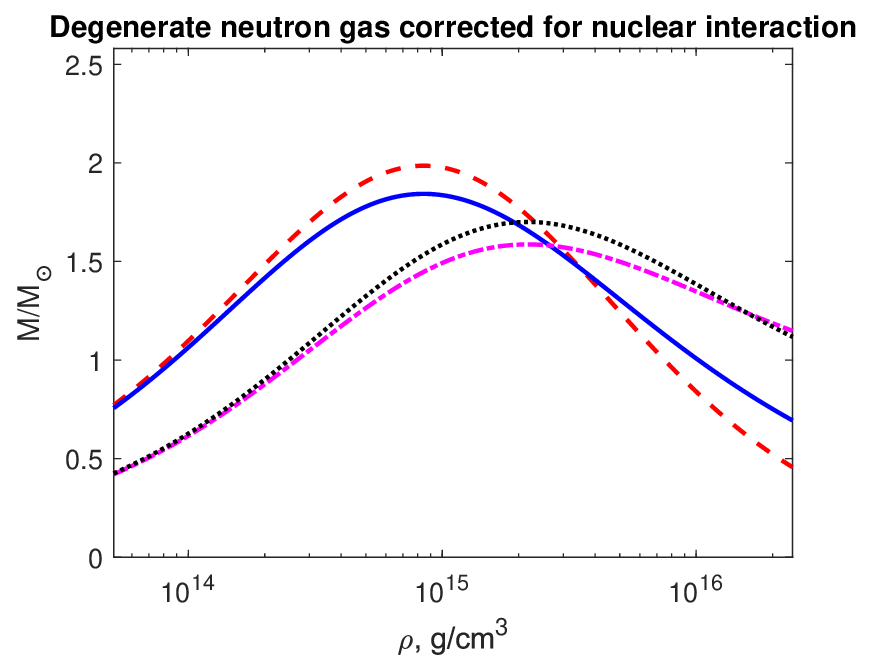}
\end{center}
\caption{Degenerate neutron gas with a correction for nuclear interaction.  Dependence $M(\rho)$ for the uniform density (solid line) and for the
exact models (dash-dotted line). Dependence $M_0(\rho)$ for the uniform-density (dashed line), and for the
exact model (dotted line).}
\label{3}
\end{figure}

\begin{figure}[!h]
\begin{center}
\includegraphics[width=0.65\linewidth]{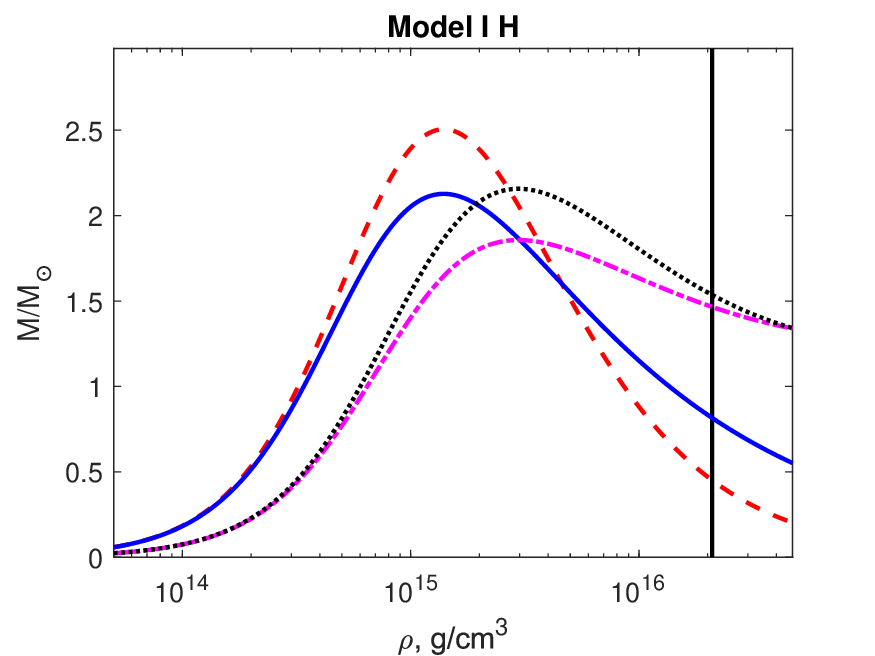}
\end{center}
\caption{ Model I H. Dependence $M(\rho)$ for the uniform density (solid line) and for the exact models (dash-dotted line). Dependence $M_0(\rho)$ for the uniform-density (dashed line), and for the exact model (dotted line).  The vertical solid line indicates the density at which $v_s=c$ and $\rho = 2.2 \times 10^{16}$ g/cm$^3$, $v_s$ is a sound speed, see \cite{zeld62}.\color {black}}
\label{4}
\end{figure}

\smallskip
\noindent
The results of calculations for several equations of state are presented in Figs.~\ref{2}–\ref{4} from \cite{bkp23}. These figures also include the dependence $M(\rho)$ for arbitrary density distributions, obtained by solving the Oppenheimer–Volkoff equations \cite{oveq}.

A comparison of uniform-density models with exact solutions of the equilibrium differential equations for various equations of state, including those from Bethe et al.~(1975) \cite{bethe}, shows that the approximate value of the critical mass exceeds the exact value by no more than $\sim 20\%$. The critical densities obtained in the uniform-density models are significantly lower than the central densities in numerical models and are comparable to the average densities of those models.

For reviews of the  newer equations of state see Refs. \cite{part,oertel, burgio, senger}

The neutron star masses had been measured in numerous observations, showing a drastic difference in errors of different measurements.

In the observations of 2 binary millisecond pulsars  NS masses are determined with unprecedented exactness due to measurements of GR effects. In one of them masses are 1.338 and 1.249 solar mass with the error box 0.001 \cite{kramer}. Measurements of NS masses of ms PSR J0348+0432 in binariy have been done, using  radio NS, and optical observations of the  companion (low mass WD) \cite{afw2013}. The authors claim  the result
 $M=  1.90 – 2.18 M_\odot$, at $3\sigma$ level.
    The high precision of this result is probably connected with using of too high precision for the companion white dwarf mass $M_{wd}=0.172\pm 0,003 \,\, M_\odot$. 
  Mass of NS in the binary with 
  ms PSR J1614-2230 was obtained by measurement of GR
Sharipo   effect from radio observations \cite{dpr2010}. 
The  authors claim the result of $M=1.85-2.09  M_\odot$, at $3\sigma$ level. In obtaining this result the authors have used formulae derived in \cite{dd1986} for two compact objects NS and BH. The companion of ms pulsar in this system is white dwarf, which tidal deformation are important, and decline the precision of the results. 
Constrains to EoS and NS parameters  have been obtained  
from  NICER's observation in combination with multimessenger observations, of the PSR J0740+6620 in \cite
{rgh2021}, and of
the PSR J0030+0451 in \cite{rwb2019}.
The precise radio measurements of the
 ms radio pulsar PSR J0437–4715  
had been described in \cite{rb2024}. Using 
relativistic corrections from \cite{dd1986},
the authors obtain the NS mass as 
$M=1,418 \pm 0.044 M_\odot$. 
The possible influence of tidal corrections from the 
helium white dwarf companion had not been discussed.

Earlier observation for NS mass measurement gave the 
following resurts for high mass NS cases.
The mass of the neutron star in Vela X-1 obtained from 
optical observations, is obtained 
\cite{bp2001} as $M=1.86 \pm 0.16 M_\odot$  (here and 
below the errors are given for $1\sigma$ confidence 
level). 
The mass and the radius of the neutron star in the 
transient low-mass X-Ray Binary SAX J1748.9-2021 were
determined in \cite{go2013} on the base of the 
observations on the X-ray satellite RXTE. One of the two 
answers, after data development, gave the value of NS 
mass $M_{NS} = 1.78 \pm 0.3 M_\odot$. 
Neutron star mass and radius measurements from 
atmospheric model fits to X-ray burst cooling tail 
spectra 4U 1702 - 429, obtained by  RXTE, 
gave the result for NS mass $M= 1.9 \pm 0.3$ 
\cite{nm2017}.  
The review of measured NS properties is given also in 
\cite{of2016}.

Of the order of 100 sources of different nature, containing neutron stars, have been collected together in \cite{fhj2024}.  Distribution of NS number as function of estimated NS masses have been plotted in one figure. From this figure the authors extract maximum mass of NS equal to 2.25 $M_\odot$, with about 3\% precision. In spite of  big statistical  material, this result cannot be fully accepted, because it does not take into account  systematic errors, which could be much higher.

Our results have similar qualitative character as  results of integration of  Oppenheimer-Volkoff  differential equation \cite{oveq}. They could be interesting for giving considerable simplification  for development of observational data. The uniform model permits to obtain rapidly connections between integral parameters of NS for different models of nuclear interaction, what, in combination with observational data,  could give  the information for improvement our knowledge about  properties of nuclear matter at densities, exceeding the density in the atomic nuclei.   
Some constrains to the EoS have been from  the registration of GW signal from two merging NS \cite{eta2018}. 

\color{black}

\section{Uniform-density strange stars}

In \cite{wit}, it was suggested that strange matter containing the strange $s$-quark may have zero pressure at a density close to nuclear density. This hypothesis was formalized within the MIT bag model, in which the corresponding equation of state is written as \cite{wit}

\begin{equation}
\label{eqs1} 
P=\frac{\rho c^2 - 4B}{3}, \quad \rho c^2 = \frac{4B}{1 - 3Q(x)}
\end{equation}
where the parameter $B$ defines the density at which the pressure vanishes, $\rho_0 = 4B/c^2$.

The solution of the algebraic equilibrium equation for this equation of state can be obtained by substituting equation~\eqref{eqs1} into equation~\eqref{eq7}, and can be written in the form \cite{bkp26}

\begin{equation}
\label{eqs3} 
M= \frac{1}{4}\big(\frac{3}{2\pi}\big)^{1/2}\frac{c^4}{G^{3/2}\sqrt{4B}} x^3[1-3Q(x)]^{1/2},\quad  
R=\frac{1}{2}\big(\frac{3}{2\pi}\big)^{1/2} \frac{c^2}{\sqrt{4GB}}x[1-3Q(x)]^{1/2},
\end{equation}
where
\[
\frac{P}{\rho c^2} = \frac{\Phi_0(x)}{\Phi_1(x)} = Q(x),
\]
	and the functions $\Phi_0(x)$ and $\Phi_1(x)$ are defined by equation~(8).

Using equation~\eqref{eqs1}, we obtain the dependences $M(\rho)$ and $M(R)$ for homogeneous strange stars, presented in Figs.~5 and 6 from \cite{bkp26}.

\begin{figure}
\centering
\includegraphics[width=0.7\linewidth]{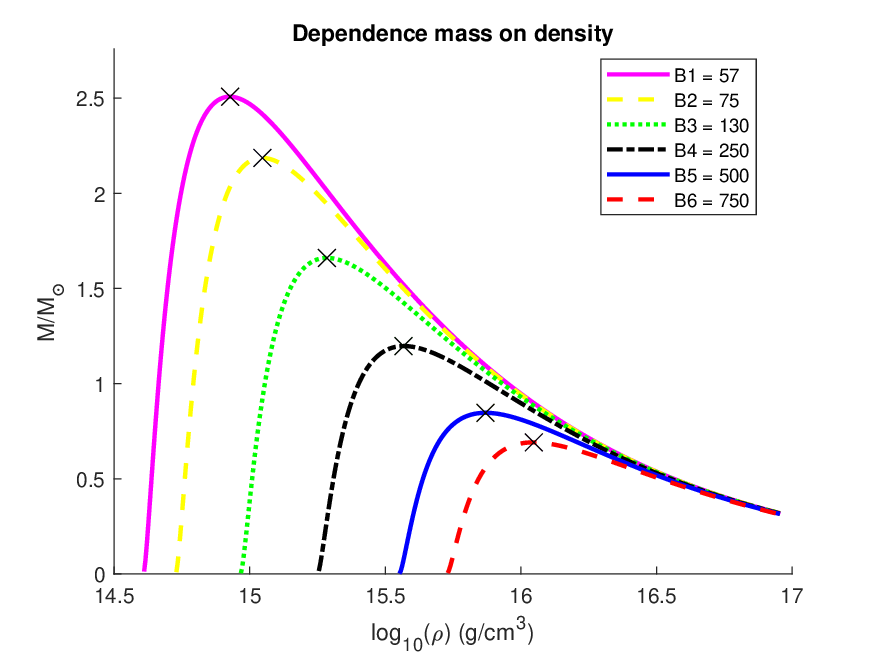}
\caption{Dependence of mass on density $M(\rho)$ for different values of the bag constant $B$ in the MIT bag model. Here $B$ is given in units of MeV/fm$^3$. The curve maxima are indicated by crosses. The parameters of stars at the maxima corresponding to critical states are presented in Table~1.}
\label{fig:fig1}
\end{figure}

\begin{figure}
\centering
\includegraphics[width=0.7\linewidth]{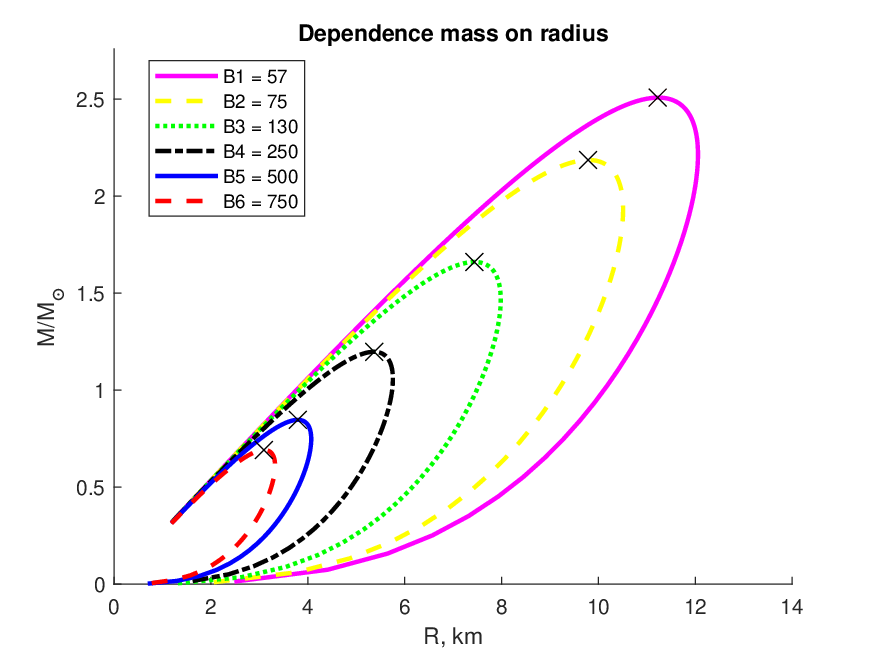}
\caption{Dependence of mass on radius $M(R)$ for different values of the bag constant $B$ in the MIT bag model. Here $B$ is given in units of MeV/fm$^3$. The curve maxima are indicated by crosses. The parameters of stars at the maxima corresponding to critical states are presented in Table~1.}
\label{fig:fig2}
\end{figure}

\noindent
The equilibrium state follows from the equation
\begin{equation}
\frac{dM}{d\rho} = \frac{dM}{dx}\frac{dx}{d\rho} = 0,
\end{equation}
which reduces to finding the extremum of the function
\begin{equation}
\Psi_m = x^3\sqrt{1 - 3Q(x)}.
\end{equation}
This extremum is determined by the equation
\begin{equation}
\frac{d\Psi_m}{dx}
= \frac{3x^2}{2\sqrt{1-3Q(x)}}\left[2(1-3Q(x)) - xQ'(x)\right] = 0.
\end{equation}
The numerical solution of this equation yields $x_m = 0.815$, which is valid for all values of the parameter $B$ in the equation of state. This value determines the critical parameters $M_m$, $\rho_m$, and $R_m$ as functions of $B$. In all models, the loss of stability occurs at the ratio $R_m/R_g = 1/x_m^2 = 1.5055$, independent of the value of $B$. The dependence of the function $\Psi_m(x)$ is shown in Fig.~\ref{5}.
Critical parameters as function of $B$ can be written in the form:
\begin{eqnarray}
\label{eqs4} 
M_m= \frac{1}{4}\big(\frac{3}{2\pi}\big)^{1/2}\frac{c^4}{G^{3/2}\sqrt{4B}} x_m^3[1-3Q(x_m)]^{1/2}=\frac{1.52\cdot 10^{18}}{\sqrt{4B}}\, M_\odot, \qquad\\
R_m=\frac{1}{2}\big(\frac{3}{2\pi}\big)^{1/2} \frac{c^2}{\sqrt{4GB}}x_m[1-3Q(x_m)]^{1/2}=\frac{6.78 \cdot 10^{23}}{\sqrt{4B}}\, {\mbox{cm}} \qquad\\
\rho_m = \frac{4B}{c^2(1-3Q(x_m))}=1.65 \cdot 10^{-21} 4B\, g/cm^3 = 8.36 \frac{B}{c^2},
\qquad x_m = 0.815,\,\,\, Q(x_m)= 0.174
\end{eqnarray}
Critical parameters of SS in the case of uniform density distribution are given in Table 1.
\begin{table}[!ht]
 \centering
 \caption{Critical parameters of homogeneous strange stars for different $B$. The density of the deconfinement point is defined by equation $\rho_0 = 4B/c^2$ g/cm$^3$.}
\begin{tabular}{|l|l|l|l|l|}
\hline
B, MeV/fm$^3$ & $\rho_0 / 10^{15}$, g/cm$^3$   & $M/M_{\odot}$    & $\rho/10^{15}$, g/cm$^3$ & R, km     \\ \hline
57  & 0.4 &2.51 & 0.839          & 11.25 \\ \hline
75  &  0.53 &2.19 & 1.106          & 9.81  \\ \hline
130 & 0.93 &1.66 & 1.919          & 7.45  \\ \hline
250 &  1.78 &1.2  & 3.703          & 5.36  \\ \hline
500 & 3.56 &0.85 & 7.372          & 3.8   \\ \hline
750 & 5.35 &0.69 & 11.15          & 3.09  \\ \hline
\end{tabular}
\end{table}

\noindent Particle physicists use the system of units, where energy density of $B$ is written in the system,  where the energy is in Mev=$1.6 10^{-6}$ erg, and the length is in fm=$10^{-13}$ cm. The astrophysical parameters are measured in CGS (cm, gram, sec) units, so we need to express the energy density $B$, known in (MeV/fm$^3$),  to its value in CGS units (erg/cm$^3$).  To obtain all parameters in CGS, it is required to use $B$ (erg/cm$^3$) = 1.78 $\cdot 10^{12}$ 
B(MeV/fm$^3$) c$^2$.

\begin{figure} 
     \centering
     \includegraphics[width=0.7\linewidth]{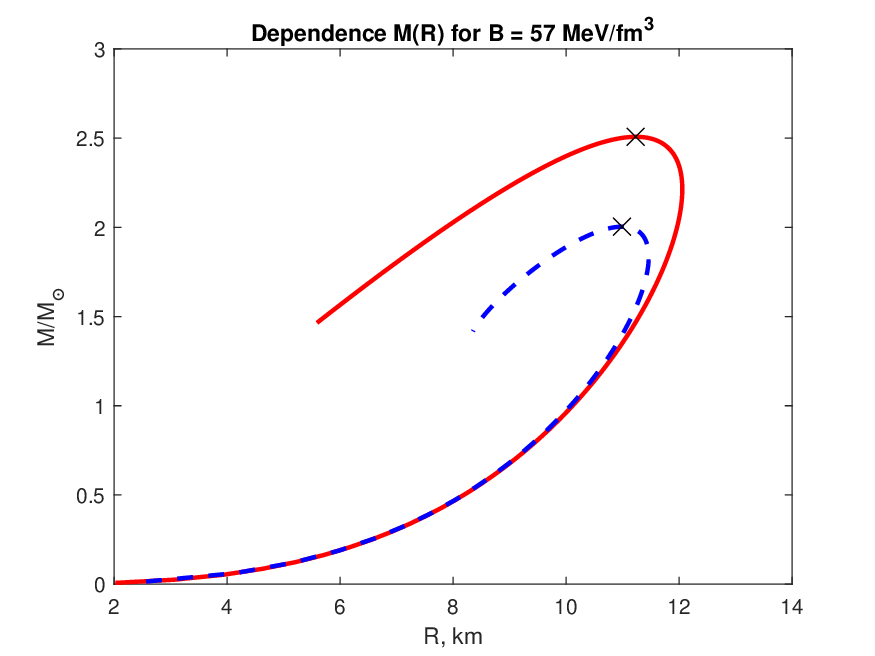}
     \caption{ Dependence  $M(R)$ for $B = 57$ MeV/fm$^3$. The solid red line shows the dependence for the uniform model,  the dashed blue line shows the exact model. The curve maxima are indicated by crosses.}
  \label{8}   
 \end{figure}

\begin{figure}
     \centering
     \includegraphics[width=0.7\linewidth]{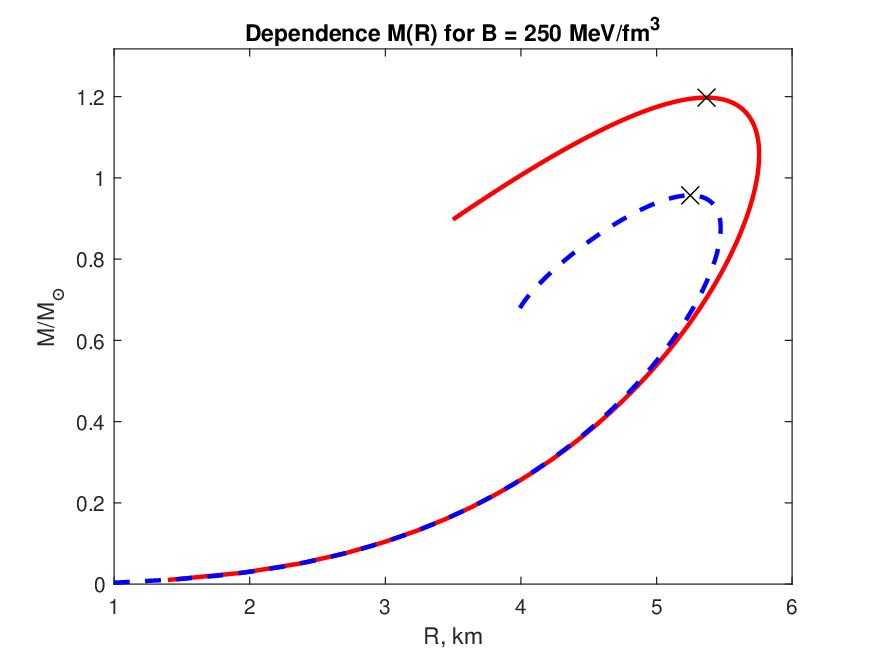}
     \caption{ Dependence  $M(R)$ for $B = 250$ MeV/fm$^3$. The solid red line shows the dependence for the uniform model,  the dashed blue line shows the exact model. The curve maxima are indicated by crosses.}
 \label{9}    
 \end{figure}

\begin{figure}
     \centering
     \includegraphics[width=0.7\linewidth]{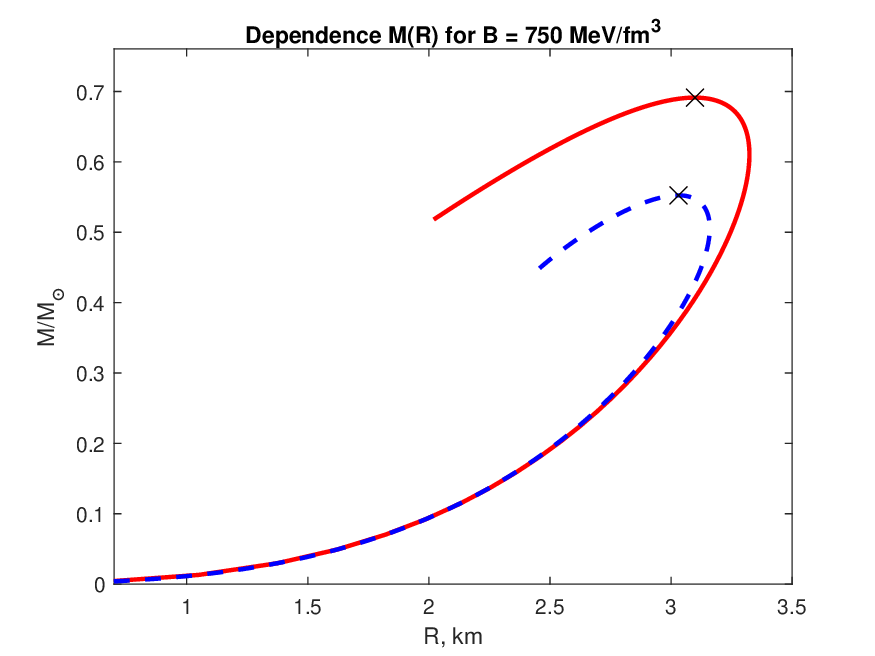}
     \caption{ Dependence  $M(R)$ for $B = 750$ MeV/fm$^3$. The solid red line shows the dependence for the uniform model,  the dashed blue line shows the exact model. The curve maxima are indicated by crosses.}
  \label{10}    
 \end{figure}

\begin{figure}
\centering
\includegraphics[width=0.7\linewidth]{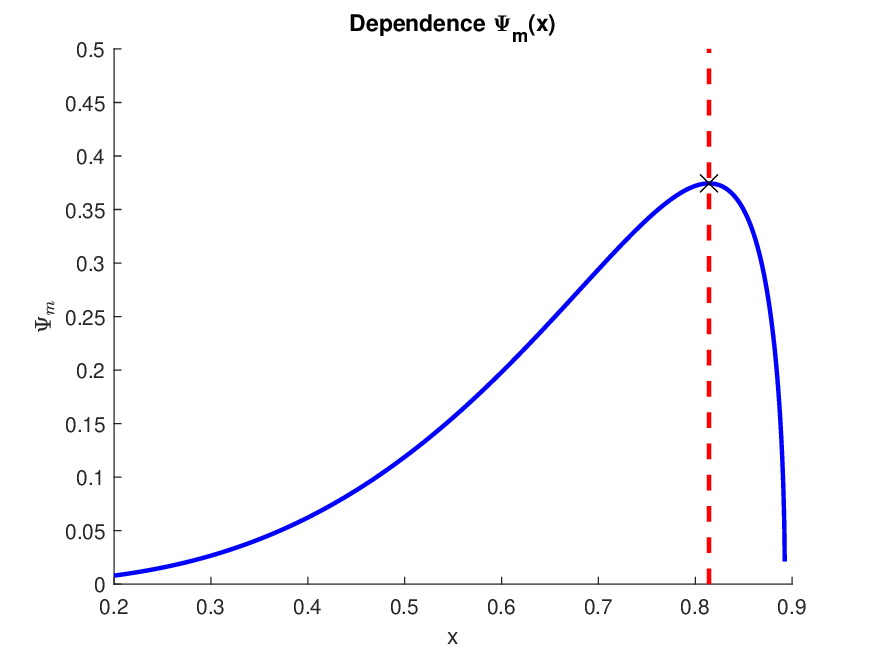}
\caption{Dependence $\Psi_m(x)$. The maximum of the function is indicated by a cross, corresponding to $x_m = 0.815$.}
\label{5}
\end{figure}

For the point of maximum mass, similar dependences were obtained in \cite{wit} by numerically solving the Oppenheimer-Volkov equation \cite{oveq}, see also \cite{nch1976}. 
To compare uniform and nonuniform density models of SS, models of cold SS were constructed for the case of a nonuniform density distribution, by integration of the Oppenheimer-Volkoff equation with the MIT bag equation of state. 
Figs. \ref{8} - \ref{10} show the mass-radius dependences for models with uniform density distribution and for the exact ones for different bag constants. 
The values at which nonuniform stars lose stability are given in Table 2.
 

 \begin{table}[!ht]
 \centering
 \caption{Critical parameters of exact model for different $B$. $\rho_0 = 4B/c^2$ g/cm$^3$ is a density of the deconfinement point.}
\begin{tabular}{|l|l|l|l|l|}
\hline
B, MeV/fm$^3$ & $\rho_0 / 10^{15}$, g/cm$^3$   & $M/M_{\odot}$    & $\rho/10^{15}$, g/cm$^3$ & R, km     \\ \hline
57  & 0.4 &2 &  1.96         & 11 \\ \hline
75  &  0.53 & 1.75 & 2.58   & 9.58  \\ \hline
130 & 0.93 & 1.33 & 4.44   & 7.28  \\ \hline
250 &  1.78 & 0.96  & 8.62    & 5.25  \\ \hline
500 & 3.56 &0.68 & 17.17          & 3.71   \\ \hline
750 & 5.35 & 0.55 & 25.8          & 3.03  \\ \hline
\end{tabular}
\end{table}

More complicated, than Bag model, for the equation of state  in strange stars was developed  in \cite{lg2015}. 
\color{black}

\section{Strange stars and confinement}

The Standard Model contains six quark flavors $(q)$: up $(u)$ $m_{u} = 2.3 \pm 0.7 \,\ MeV/c^2$, down $(d)$ $m_{d} = 4.8 \pm 0.5 \,\ MeV/c^2$, strange $(s)$ $m_{s} = 95 \pm 5 \,\ MeV/c^2$, charm $(c)$ $m_{c} = 1275 \pm 25 \,\ MeV/c^2$, bottom $(b)$  $m_{b} = 4180 \pm 30 \,\ MeV/c^2$, and top $(t)$ $m_{t} = 173210 \pm 25 \,\ MeV/c^2$.  Baryons containing one or more strange quarks, but no charm $(c)$, bottom $(b)$, or top $(t)$ quarks, are called hyperons. Hyperons may exist in stable form in the cores of neutron stars.  Some types of hyperons are : Lambda $\Lambda^0$, consisting form $uds$ quarks, $m_{\Lambda^0} = 1115.68 \,\ MeV/c^2$; Sigma $\Sigma^+$, consisting from $uus$, $m_{\Sigma^+} = 1189.37 \,\ MeV/c^2$; Omega $\Omega^-$, consisting from $sss$ quarks, $m_{\Omega^-} = 1672.45 \,\ MeV/c^2$.

\subsection{Experimental estimations}

The energy density in CERN SPS head-on experiments for the transition to quark–gluon plasma (QGP) was estimated to be about $3~\mathrm{GeV/fm}^3 \approx 5.6 \times 10^{15}~\mathrm{g/cm}^3$ \cite{alber}. Experimental data from the Relativistic Heavy Ion Collider (RHIC) gave a lower limit for QGP formation of $\epsilon \approx 5~\mathrm{GeV/fm}^3 = 8.91 \times 10^{15}~\mathrm{g/cm}^3$ \cite{brahms}. Data from PHENIX (RHIC) indicate that the energy density at the point of QGP formation is at least $15~\mathrm{GeV/fm}^3 = 2.7 \times 10^{16}~\mathrm{g/cm}^3$ \cite{phenix}. In experiments at the LHC \cite{alice2016}, the central energy density during QGP formation in Pb+Pb collisions was estimated as $\epsilon \approx 21~\mathrm{GeV/fm}^3$.

\subsection{Theoretical estimations}

Numerous theoretical investigations based on simplifying assumptions have been performed over the past $\sim 30$ years.

Strange stars have been extensively studied in the particle physics, including astrophysical applications. Strange matter and quark stars have been discussed in many reviews \cite{weber}, \cite{yang}, \cite{tolos}, \cite{xia}, and some
textbooks \cite{glendenning}, \cite{haensel}.

Their predictions for the QGP formation boundary are scattered around $150$–$200~\mathrm{MeV}$, which is more than an order of magnitude lower than experimental estimates. This discrepancy suggests that further theoretical improvements are required, possibly involving more sophisticated models.

  Theoretical investigation about deconfinement  process had been done in \cite{jxs2022}. Observational search of strange stars is started soon after publication of the paper \cite{wit}. The existence of a strange star in the compact object is based on measurements of its small radius, which cannot be less $\sim 8\,M_\odot$ in the neutron star. While the mass of the compact objects  was estimated in observation of many binary sources, the measurements of radius are critically based on the model a heat transfer. The famous claim about possible discovery of a strange star appears in the paper \cite{drake}.  
 Basing on Chandra X-ray measurements of the X-ray source RX J1856.5-3754, the authors came to conclusion, that its radius R= 3.8-8.2 km is too small for the NS, indicating to the possible strange quark star.  
Subsequent calculations based on the improved heat transfer model \cite{ttb2004,tbh2004}, using the same Chandra, and XMM X-ray measurements, gave quite another result for the radius of  the compact star  $R\ge 13.3$ km, appropriate to NS. 
  The review about recent progresses in strange quark stars is given in \cite{zhz2024}.
\color{black}

\section{Conclusion}

A comparison of the parameters obtained in the calculated models with observational results for NS, and acceleration experiments shows, that the values of the bag constant $B$ can be considered realistic only for large values of $B \sim 750$, or even larger. At large $B$, the loss of stability of a strange star occurs at masses $\lesssim (0.5$–$0.7)\,M_{\odot}$, which is significantly lower than the maximum neutron star masses for all equations of state. We therefore conclude that strange stars are unlikely to form during stellar evolution.

This conclusion is based on the Bag model, which itself is purely theoretical, and model dependent on the  quantum field theory calculations.
  In absence of any reliable observational data (these objects are not yet discovered) we may deal only with theoretical instruments for EoS, in which  the bag model is the most popular. This model depends on the 
theoretical approximate  results of the relativistic quantum theory, which itself is not fully established.
Observational NS data development, using different theoretical models, permits to obtain restrictions for the choice of these models. That leads to better understanding of some nuclear physics  problems. This is not possible for quark matter, but  theoretical consideration may serve as indication to search a way, where these objects could be discovered.

The matter density in strange stars is comparable to the density at very early stages of the expansion of the Universe, where strange stars could, in principle, form with a wide range of masses, independently of stellar evolution. Such objects may be referred to as \emph{Primordial Strange Stars} (PSS), in analogy with Primordial Black Holes (PBHs), should either of these objects exist.

\end{document}